\documentclass[runningheads]{llncs}

\usepackage[T1]{fontenc}
\usepackage{xcolor}
\usepackage{listings}
\usepackage{inconsolata}
\usepackage{relsize}
\usepackage{amsmath,amssymb,array}
\usepackage{esz}
\usepackage{mathpartir}
\usepackage{graphicx}
\usepackage{tikz}
\usepackage{xspace}
\usepackage{xurl}
\usepackage{caption}
\usepackage{bbding}
\usepackage{tabularx}
\usepackage{booktabs}
\usepackage{enumitem}
\usepackage{cite}
\makeatletter
\let\titlesec@oldsubparagraph\subparagraph
\let\subparagraph\paragraph
\makeatother
\usepackage[compact]{titlesec}
\makeatletter
\let\subparagraph\titlesec@oldsubparagraph
\makeatother
\usepackage{hyperref}
\usepackage{microtype}

\hypersetup{colorlinks=true,linkcolor=blue,citecolor=blue,urlcolor=blue}

\titlespacing*{\section}{0pt}{0.9ex plus 0.15ex minus 0.15ex}{0.45ex plus 0.1ex}
\titlespacing*{\subsection}{0pt}{0.6ex plus 0.1ex minus 0.1ex}{0.3ex plus 0.1ex}
\titlespacing*{\subsubsection}{0pt}{0.5ex plus 0.1ex minus 0.1ex}{0.3ex plus 0.1ex}
\titlespacing*{\paragraph}{0pt}{0.6ex plus 0.1ex minus 0.1ex}{0.6em}

\newlist{Description}{description}{1}
\setlist[Description]{leftmargin=1em, labelindent=1em, style=unboxed, font=\bfseries, itemsep=0.25ex, topsep=0.25ex, parsep=0pt}

\newlist{Itemize}{itemize}{1}
\setlist[Itemize]{label=--, leftmargin=1em, labelindent=0pt, labelsep=0.5em, itemsep=0.25ex, topsep=0.25ex, parsep=0pt}

\newif\ifcomments
\commentsfalse

\newenvironment{Figure}{
    \begin{figure}[tb]
} {
    \end{figure}
}

\newenvironment{masm}{\zedindent=2ex\begin{zed}}{\end{zed}}

\newcommand{\MOVELOC}{\Zpreop{move\_loc}}

\newcommand{\BORROWLOC}{\Zpreop{borrow\_loc}}
\newcommand{\BORROWFIELD}{\Zpreop{borrow\_field}}

\newcommand{\sat}{\mathrel{{\scriptstyle\vert}\mkern-1mu{\scriptstyle\sim}}}

\lstdefinestyle{MoveStyle}{
    basicstyle=\ttfamily,
    keywordstyle=\color{black}, 
    commentstyle=\color{gray}\normalfont\itshape,
    escapechar=@, 
    aboveskip=0.4\baselineskip, 
    belowskip=0.4\baselineskip,
    breaklines=true,
    literate={|~}{{$\sat$\,}}1
        {<=}{{$\leq$}}2
        {>=}{{$\geq$}}2
        {==}{{$=$}}2
        {!=}{{$\neq$}}2
        {==>}{{$\Longrightarrow$}}3
        {forall}{{$\forall$}}1
        {exists}{{$\exists$}}1,
}

\lstdefinelanguage{Move}{
    morekeywords={
        abort,
        aborts_if,
        aborts_of,
        acquires,
        address,
        apply,
        as,
        assert,
        assume,
        borrow_global,
        borrow_global_mut,
        break,
        const,
        continue,
        copy,
        copyable,
        define,
        drop,
        else,
        ensures,
        ensures_of,
        enum,
        exists,
        false,
        forall,
        friend,
        fun,
        global,
        has,
        havoc,
        if,
        in,
        include,
        invariant,
        key,
        let,
        loop,
        match,
        modifies,
        modifies_of,
        module,
        move,
        move_from,
        move_to,
        mut,
        native,
        num,
        phantom,
        old,
        onabort,
        pragma,
        proof,
        public,
        requires,
        requires_of,
        resource,
        result_of,
        return,
        schema,
        script,
        signer,
        spec,
        split,
        store,
        struct,
        true,
        u8,
        u64,
        u128,
        u256,
        update,
        use,
        with,
        where,
        while},
    sensitive=true,
    morecomment=[l]{//},
    morecomment=[s]{/*}{*/},
}

\lstMakeShortInline[language=Move,style=MoveStyle,breaklines=true,breakatwhitespace=true,escapechar=@]|
\lstMakeShortInline[language=Move,style=MoveStyle,breaklines=true,breakatwhitespace=true,escapechar=@]!

\lstnewenvironment{Move}{
    \lstset{
        language=Move,
        style=MoveStyle,
        basicstyle=\relsize{-2}\ttfamily,
        keywordstyle=\color{blue},
        xleftmargin=2em,
    }
}{
}

\lstnewenvironment{MoveBox}{
    \lstset{
        language=Move,
        style=MoveStyle,
        basicstyle=\relsize{-2}\ttfamily,
        keywordstyle=\color{blue},
        xleftmargin=2em,
    }
}{
}

\lstnewenvironment{MoveBoxNumbered}{
    \lstset{
        language=Move,
        style=MoveStyle,
        basicstyle=\relsize{-2}\ttfamily,
        keywordstyle=\color{blue},
        xleftmargin=2em,
        numbers=left,
        numberstyle=\scriptsize\color{gray},
    }
}{
}

\lstnewenvironment{MoveDiag}{
    \lstset{
        style=MoveStyle,
        basicstyle=\relsize{-2}\ttfamily,
    }
}{
}

\lstnewenvironment{claude}{
    \lstset{
        basicstyle=\relsize{-2}\ttfamily,
        breaklines=true,
        breakautoindent=true,
        breakatwhitespace=true,
        breakindent=1.2em,
        columns=fullflexible,
        keepspaces=true,
        escapechar=@,
        extendedchars=true,
        inputencoding=utf8,
        literate=
            {⎿}{{$\llcorner$}}1
            {└}{{$\llcorner$}}1
            {├}{{$\vdash$}}1
            {─}{{-}}1
            {│}{{$\vert$}}1
            {◻}{{$\square$}}1
            {◼}{{$\blacksquare$}}1
            {☐}{{$\square$}}1
            {☑}{{$\boxtimes$}}1
            {●}{{$\bullet$}}1
            {○}{{$\circ$}}1
            {◯}{{$\bigcirc$}}1
            {⏺}{{$\bullet$}}1
            {✽}{{$\ast$}}1
            {★}{{$\star$}}1
            {☆}{{$\star$}}1
            {…}{{\ldots}}1
            {↗}{{$\nearrow$}}1
            {↘}{{$\searrow$}}1
            {↳}{{$\hookrightarrow$}}1
            {✓}{{\Checkmark}}1
            {✗}{{\XSolidBrush}}1
            {•}{{$\bullet$}}1
            {–}{{--}}1
            {—}{{---}}1
            {‘}{{`}}1
            {’}{{'}}1
            {“}{{``}}1
            {”}{{''}}1,
        moredelim=[l][\color{blue}\bfseries]{>},
    }
}{
}

\begin{document}

\title{Defense-in-Depth Runtime Safety in Move}
\titlerunning{Defense-in-Depth Runtime Safety in Move}

\author{Victor Gao \and
        Wolfgang Grieskamp\thanks{Corresponding authors: Wolfgang Grieskamp (\email{wg@aptoslabs.com}) and Vineeth Kashyap (\email{vineeth@aptoslabs.com}).} \and
        Vineeth Kashyap\textsuperscript{$\star$} \and
        George Mitenkov \and \\
        Teng Zhang \and
        Runtian Zhou \and
        Andrea Cappa \and
        Marco Ilardi}
\authorrunning{V. Gao et al.}
\institute{Aptos Labs, Palo Alto, USA}

\maketitle

\begin{abstract}
Move is a smart-contract language used to execute transactions on the Aptos blockchain.
Move programs execute in a sandboxed VM as typed bytecode.
The VM statically verifies foundational safety properties like type safety and reference safety at code loading time.
In principle, this design gives strong guarantees for Move.
However, the static verification logic is complex and continually evolving with the language; like any software, it is not immune to bugs.
In a live blockchain setting, a missed rule violation can translate directly into loss of assets, forged authority, or unrecoverable corruption of on-chain state.
For this reason, Aptos relies on defense-in-depth \emph{runtime safety checks} that independently verify the critical invariants during execution, providing protection against latent verifier bugs and malicious bytecode.
This paper motivates and describes the runtime safety checks for Move on Aptos.
\end{abstract}

\section{Introduction}
\label{sec:intro}

The Move on Aptos language~\cite{MoveOnAptosBook} was originally developed for the Libra/Diem blockchain at Meta~\cite{libra_blockchain_white,OLD_MOVE_LANG,blackshear2020resources} and is now deployed at scale on Aptos~\cite{Aptos}, a high-throughput Layer-1 blockchain that settles transactions for billions of dollars in user assets on chain.
A defining trait of the language is its resource-oriented type discipline: digital assets are represented as Move \emph{resources} whose ownership and movement are tracked by linear-style abilities (|copy|, |drop|, |store|, |key|), ruling out accidental duplication or destruction at compile time, and supported by a regime of mutable and immutable references similar to the Rust language.
The Aptos platform builds on this foundation with a system of objects, parallel execution, and a rich on-chain framework written in Move itself~\cite{AptosFrameworkBook,ParkZGXGCLC24}.

Move has been designed for \emph{offline formal verification}: the Move Prover translates Move programs and their specifications, written in the Move Specification Language, into verification conditions discharged by SMT solvers~\cite{DillGPQXZ22,GrieskampEtAl26HO}.
The Move Prover can verify rich properties about the smart contracts themselves, but it is inherently offline, applied prior to deployment of the contracts.

Move programs are compiled into typed bytecode executed by the Move VM.
The VM performs \emph{static bytecode verification} when a module is loaded, enforcing well-formedness, typing, and Move's reference (\S\ref{sec:move}) discipline.
The reference safety verifier~\cite{MoveBorrowChecker} is particularly intricate: it uses abstract interpretation to maintain a borrow graph that conservatively approximates the aliasing relationships entailed by control- and data-flow.
Keeping these analyses correct is delicate, especially as the language evolves: since the public launch of Aptos, Move has gained enums, closures, and public structs, all of which interact non-trivially with the static verifier.
Given the high stakes, Aptos runs a bug bounty program, which has surfaced verifier bugs in the past.

In a permissionless blockchain setting, executed transactions cannot be reversed, and anyone can deploy hand-crafted compiler-bypassed bytecode.
Static verification is thus a single point of failure: one exploited bug can have catastrophic consequences.
As a practical security measure, the Move VM implements defense-in-depth \emph{runtime safety checks} that, after static verification, independently verify the critical invariants during execution.
These runtime safety checks do not share implementation with the static verifier.
Multiple proof-of-concept exploits, surfaced through bug bounty reports and internal audits, bypassed the static verifier but were caught by the runtime checks: the defense in depth working as intended.

In this paper, we describe three groups of runtime checks, each motivated by real-world examples.
\emph{Type checks} (\S\ref{sec:type}) verify that the dynamically tracked type of every value matches the static type the bytecode expects to consume.
\emph{Ability checks} (\S\ref{sec:ability}) verify that operations such as copy, drop, or store to the global state are only performed on values whose declared abilities permit them.
\emph{Reference checks} (\S\ref{sec:ref}) verify the same freedom from dangling references and aliased mutable access that the static verifier is supposed to enforce, but using fundamentally different representations and algorithms.

We then describe two techniques for reducing the runtime costs of these checks: \emph{trusted code} (\S\ref{sec:trusted}) that waives them under certain conditions, and an \emph{asynchronous mode} (\S\ref{sec:async}) that leverages idle threads in Block-STM~\cite{BlockSTM} (Aptos's parallel transaction execution engine).
We present an evaluation of the performance impact of the checks on real-world benchmarks (\S\ref{sec:perf}).
Our implementations are openly available~\cite{AptosCore}.

\section{Move on Aptos}
\label{sec:move}

We briefly summarize key aspects of Move on Aptos~\cite{MoveOnAptosBook,blackshear2020resources}.

\paragraph{Typing and generics.}
Move is strongly statically typed.
Generic functions and types are instantiated by substitution at the call site.
Substitution is carried through transitively: accesses to type-indexed global storage (below) are keyed by concrete types at runtime.

\paragraph{Abilities.}
Each type carries a subset of the four abilities: |copy|, |drop|, |store|, and |key|.
Types without |drop| or |copy| are linear: every value of such a type must be moved, globally stored, or explicitly destructured.
|key| indicates a type which can be stored as a global resource, and |store| a type which can be transitively included in resources.

\paragraph{Algebraic data types.}
Structs are nominal product types; enums are nominal sums whose variants carry payloads.
Pattern matching may bind references directly into a variant's fields.

\paragraph{Modular encapsulation.}
Privileged operations (construction, destruction, and field access) on a private type are confined to its defining module, so module-external code can manipulate values only through module-provided functions.

\paragraph{References.}
Immutable (|&T|) and mutable (|&mut T|) references obey a static aliasing discipline comparable to Rust's~\cite{rust,rustbelt}: two live mutable references cannot point to overlapping memory, and references cannot outlive the values they borrow.
In contrast to Rust, Move (a) does allow mutable references to global storage, but (b) has no reference lifetimes and does not allow references to be stored in structs or enums or captured by closures.

\paragraph{Global storage.}
Global state is a map indexed by (resource type, address) pairs.
Resources are types with the |key| ability.
They are accessed through |T[addr]|, |&T[addr]|, and |&mut T[addr]|.


\section{Runtime Safety Checks}
\label{sec:bugs}

This section describes the defense-in-depth runtime checks.
Each is motivated by real-world examples of security bugs they are designed to prevent.

\subsection{Type Safety}
\label{sec:type}

Move has nominal typing: two structs with identical layout but different names are distinct types.
A \emph{type-confusion} exploit breaks that distinction, and can be catastrophic when one of the two types is privileged.
Consider the example:

\noindent
\begin{minipage}[t]{0.53\linewidth}
\begin{lstlisting}[language=Move,style=MoveStyle,basicstyle=\relsize{-2}\ttfamily,keywordstyle=\color{blue},frame=single,frameround=tttt,framesep=4pt,rulecolor=\color{gray!60},backgroundcolor=\color{gray!4}]
module 0x1::admin {
 struct Capability(address);
 public fun admin_action(c: &Capability) {
  /* Capability gated action */
 }
}
\end{lstlisting}
\end{minipage}\hfill
\begin{minipage}[t]{0.43\linewidth}
\begin{lstlisting}[language=Move,style=MoveStyle,basicstyle=\relsize{-2}\ttfamily,keywordstyle=\color{blue},frame=single,frameround=tttt,framesep=4pt,rulecolor=\color{gray!60},backgroundcolor=\color{gray!4}]
module 0xdeadbeef::attacker {
 struct FakedCapability(address);
}
\end{lstlisting}
\end{minipage}

\noindent The attacker can freely construct their owned type |FakedCapability|; a type confusion exploit would allow passing it to |admin_action| as if it were a |Capability|.

\paragraph{Example.}
When vector operations were added to the VM during Move's original development at Meta, the bytecode verifier missed an element-type check for push.
The compiler performed this check; no input source program could reach the bug.
But an attacker could directly construct bytecode, conceptually doing:

\noindent\hfil\begin{minipage}{0.78\linewidth}
\begin{lstlisting}[language=Move,style=MoveStyle,basicstyle=\relsize{-2}\ttfamily,keywordstyle=\color{blue},frame=single,frameround=tttt,framesep=4pt,rulecolor=\color{gray!60},backgroundcolor=\color{gray!4}]
let v: vector<Capability> = vector::empty();
v.push_back(FakedCapability(victim)); // should be rejected
admin_action(&v[0]);  // runs on a faked type
\end{lstlisting}
\end{minipage}\hfil

\noindent The two types share a layout, so the program runs without corruption, but the gated action runs on a faked type.
This bug was found as part of the Aptos bug bounty program before going into production.

\paragraph{Performing the Checks.}
To prevent type confusion, in addition to the bytecode verifier, type checks are performed at runtime.
A type stack is maintained shadowing the operand stack of the VM.
The VM's locals are typed, so every value's type is known. 
When a value is pushed on the operand stack, its type is pushed on the type stack.
When a value is popped, by being stored into a local or used by an operation, its type is popped and compared with the expected type of the local or operation; a mismatch is type confusion and raises a runtime error.



\subsection{Ability Safety}
\label{sec:ability}

Bugs not enforcing |copy| ability are very damaging.
A value whose type lacks |copy| is meant to be unique: this is what makes the |Capability| of \S\ref{sec:type} a single-issue permission, and what makes a |Coin| balance impossible to counterfeit.
A runtime that produces a fresh value of a no-|copy| type silently duplicates that resource.

\paragraph{Example.}
Move's ability system permits an assignment to downgrade a value's abilities (i.e., width subtyping over the ability set, which is sound).
But due to the classic invariance of mutable references~\cite{pierce2002tapl}, applying this rule through a mutable reference is unsound.
When Move was extended with closures, a corner case (see example below) did not enforce invariance.
An exploit could capture a no-|copy| resource |r| inside a no-|copy| closure |h|, store that closure into a variable |f| of |copy|-able closure type, and then copy |f| freely, calling it multiple times, duplicating the no-|copy| resource |r|.

\noindent
\begin{minipage}[t]{0.46\linewidth}
\begin{lstlisting}[language=Move,style=MoveStyle,basicstyle=\relsize{-2}\ttfamily,keywordstyle=\color{blue},frame=single,frameround=tttt,framesep=4pt,rulecolor=\color{gray!60},backgroundcolor=\color{gray!4}]
struct NoCopy(u64) has drop;

public fun assign<T: drop>(
    ref: &mut T, x: T
) {
 // old ref value dropped
 *ref = x;
}
\end{lstlisting}
\end{minipage}\hfill
\begin{minipage}[t]{0.50\linewidth}
\begin{lstlisting}[language=Move,style=MoveStyle,basicstyle=\relsize{-2}\ttfamily,keywordstyle=\color{blue},frame=single,frameround=tttt,framesep=4pt,rulecolor=\color{gray!60},backgroundcolor=\color{gray!4}]
let r = NoCopy(42);
let h: || has drop = || {
    // capture and use `r`
};
let f: || has drop + copy = || {};
assign<|| has drop>(&mut f, h);
// ^ bug if not rejected: covariant
let g = copy f; f(); g();
\end{lstlisting}
\end{minipage}

\noindent This bug was found as part of pre-production security auditing.

\paragraph{Performing the Checks.}

The VM performs runtime checks for violations of each of the abilities |copy|, |drop|, |key|, and |store|.
These checks are performed when the corresponding instruction is executed.
For |drop|, the VM tracks which values have not been consumed on function exit.
The runtime checks for |drop| are more precise than the static checks, as the latter conservatively rejects if a no-|drop| local may not be consumed on some control-flow path (including infeasible paths), while the former only checks the real executed path.


\subsection{Reference Safety}
\label{sec:ref}

The mutable-aliasing rule is critical in the presence of |enum| types: a reference into a variant payload is type-safe only while the enum still holds that variant.

\paragraph{Example.}
If two mutable references to the same enum can simultaneously be live, one can switch the variant under the other.
Consider the following code:

\noindent\hfil\begin{minipage}{0.78\linewidth}
\begin{lstlisting}[language=Move,style=MoveStyle,basicstyle=\relsize{-2}\ttfamily,keywordstyle=\color{blue},frame=single,frameround=tttt,framesep=4pt,rulecolor=\color{gray!60},backgroundcolor=\color{gray!4}]
enum E { V1(Capability), V2(FakedCapability) }

fun aliased_swap(a: &mut E, b: &mut E, faked: &mut E) {
    // faked is V2(FakedCapability(victim))
    match (a) {
        E::V1(c) => {
            mem::swap(b, faked); // *a is now V2(..)
            admin_action(c); // c still &Capability
        },
        _ => {},
    }
}
\end{lstlisting}
\end{minipage}\hfil

\noindent A call of the form |aliased_swap(p, q, ...)| where |p| and |q| alias should normally be rejected by the bytecode verifier.
However, the implementation of the static borrow analysis had a bug in its \emph{borrow graph} representation.
The analysis treats borrows along edges with distinct labels as disjoint, but the representation of labels for enum fields was not canonical: two distinct labels could denote the same field offset, though only in carefully hand-crafted bytecode (not shown here), never in compiler-emitted code.
Verification thus admitted such bytecode where |aliased_swap| could be called with aliased mutable references, resulting in the type confusion vulnerability shown in the listing above.
This bug was discovered via the bug bounty program after the introduction of enum types and was never exploited on the blockchain.

\paragraph{Performing the Checks.}
Checking reference safety at runtime for Move-style borrow semantics is novel (\S\ref{sec:concl}).
A challenge is that when constructing references, mutable reference aliasing \emph{is} allowed.
Only once those references are used must aliasing be forbidden.
For example, consider the valid code:

\noindent\hfil\begin{minipage}{0.78\linewidth}
\begin{lstlisting}[language=Move,style=MoveStyle,basicstyle=\relsize{-2}\ttfamily,keywordstyle=\color{blue},frame=single,frameround=tttt,framesep=4pt,rulecolor=\color{gray!60},backgroundcolor=\color{gray!4}]
struct S {a: u64, b: u64}
let s: S;
let r1 = &mut s.a; let r2 = &mut s.b;
*r1 = 1; *r2 = 2; // r1 & r2 reference disjoint fields
\end{lstlisting}
\end{minipage}\hfil

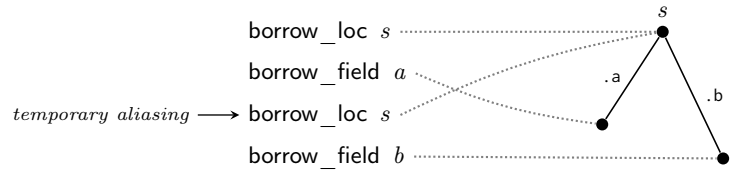
\begin{figure}[t]
\centering
\begin{tikzpicture}[
  >=stealth,
  op/.style={anchor=west, font=\small},
  dot/.style={circle, fill=black, inner sep=1.6pt},
  tedge/.style={black, line width=0.6pt},
  conn/.style={densely dotted, gray, line width=0.8pt},
]
\node[op] (i1) at (0, 0.825)  {$\BORROWLOC\ s$};
\node[op] (i2) at (0, 0.275)  {$\BORROWFIELD\ a$};
\node[op] (i3) at (0,-0.275)  {$\BORROWLOC\ s$};
\node[op] (i4) at (0,-0.825)  {$\BORROWFIELD\ b$};
\node[anchor=east, font=\scriptsize\itshape] (ta) at (-0.55,-0.275) {temporary aliasing};
\draw[->, >=stealth, line width=0.5pt] (ta.east) -- (i3.west);
\node[dot, label={[font=\small]above:$s$}] (root) at (5.6, 0.825) {};
\node[dot] (na) at (4.8,-0.4) {};
\node[dot] (nb) at (6.4,-0.85) {};
\draw[tedge] (root) -- (na) node[midway, left=1pt, font=\scriptsize] {\texttt{.a}};
\draw[tedge] (root) -- (nb) node[midway, right=1pt, font=\scriptsize] {\texttt{.b}};
\draw[conn] (i1.east) -- (root);
\draw[conn] (i2.east) to[bend right=8] (na);
\draw[conn] (i3.east) to[bend left=8] (root);
\draw[conn] (i4.east) -- (nb);
\end{tikzpicture}
\caption{A dotted line links each bytecode operation (left) to the access-path-tree node (right) for the place its pushed reference denotes.
}
\label{fig:aliastree}
\end{figure}

\noindent The corresponding stack-machine bytecode fragment is shown in Fig.~\ref{fig:aliastree}.
\noindent While we have a mutable reference to $s.a$ on the stack, we temporarily create a new mutable reference to the whole of $s$, overlapping with the existing one~(Fig.~\ref{fig:aliastree}).
The aliasing is eliminated only when this reference is consumed to create the reference to $s.b$.
Thus, a naive approach to runtime reference safety checking, for example via reference counting, would be infeasibly strict.

Our checker instead executes an \emph{abstract shadow semantics} in lockstep with execution.
The VM's operand stack, locals, and call frames are mirrored by a shadow state in which every non-reference value is opaque and every reference is a fresh identifier $\rho$ minted at its borrow site.
The shadow state never inspects runtime values or memory addresses; conceptually, it is an abstract interpretation of the single path taken, embedded along with the concrete execution.

\paragraph{State representation.} References point into trees of \emph{places}, not at each other.
Each function lazily grows \emph{access path trees} rooted at its locals, at the global resource types it borrows, and at the values behind its reference parameters.
Edges are field offsets: fields of different enum variants that share an offset share an edge.
A reference $\rho$ is a pointer to a node (representing a place) in one of these trees, tagged with its mutability.
Every aliasing question reduces to an ancestor/descendant relation between the places two references denote; no borrow genealogy among the references themselves is recorded.

\paragraph{Poisoning, lazy fails.}
A destructive operation, like a write through a mutable reference or the invalidation of a local by $\MOVELOC$, poisons every other reference whose places it invalidates.
No error is raised at the operation itself.
If a poisoned reference is later \emph{used} --- read, written, borrowed through, or passed to a callee --- a transaction-aborting error is raised.
Poisoned references may be moved around or dropped without error.
Overlapping references that never destructively interact are never flagged at all, admitting the temporary aliasing illustrated above.

\paragraph{Compositional.} The checker is compositional across function boundaries.
A call consumes its reference arguments, checks at the call site that the places reachable from each mutable argument are disjoint from those reachable from every other reference argument, and hands the callee fresh trees rooted at its own reference parameters.
Passing a mutable reference into a call is a destructive operation, to keep the checker compositional (callee may mutate arbitrarily through it).
On return, every returned reference must denote a place inside the tree of some reference parameter; its access path is replayed into the corresponding tree of the caller.
Returned mutable references must also denote mutually disjoint places.
Hence, the checker tracks precisely which parameter each returned reference derives from, along the executed path, while the static verifier conservatively assumes that a returned reference may borrow from every reference argument.

Thus, the checker realizes a \emph{relaxed dynamic semantics} of the static borrow analysis: (a) bytecode accepted by a correct static verifier is expected to pass the runtime checks on every execution, and (b) the checker admits strictly more behaviors, because it observes only the executed path and has none of the verifier's join and call-summary over-approximations.
We have validated (a) empirically --- through extensive testing, fuzzing, and re-execution of billions of historical Aptos transactions --- without observing a false positive.
We have confirmed (b) by hand-crafting bytecode that the static verifier rejects, yet which executes without violating reference safety and is hence admitted by the checker.

For the example (Fig.~\ref{fig:aliastree}), the bytecode fragment runs successfully: the second $\BORROWLOC\ s$ briefly creates a second mutable reference at the tree root for $s$, but that reference is consumed by $\BORROWFIELD\ b$ before any destructive write fires, so nothing is poisoned.

The exploit described earlier in this section, that creates the aliased mutable references, is caught by poisoning instead: edges in the enum's access path tree are field offsets, so the two references land on the same node by construction.
The destructive |mem::swap| then poisons the still-live payload reference |c| sitting at a descendant place, and the |admin_action(c)| aborts because it is |c|'s use.




\section{Trusted Code}
\label{sec:trusted}

Because runtime checks impose a performance overhead, the VM supports \emph{trusted code}: modules for which these checks are waived.
Trusted deployment is currently under Aptos Foundation governance and limited to the Aptos framework; each deployment is approved by the Aptos operators, and we are also investigating staking-based models.
Unlike arbitrary bytecode published on-chain, trusted code is always compiled from source, and thus passes the compiler's extensive type, ability, and reference safety checks before undergoing static bytecode verification.
Defense in depth is therefore preserved: for trusted code, the compiler and the static verifier form the two independent layers, with the compiler playing the role that the runtime checks play for untrusted code.

Executing trusted code disables the checks of \S\ref{sec:type} and \S\ref{sec:ability}.
Because checks like type safety accumulate state during execution, the transitions between untrusted and trusted code must be modeled carefully.
For example, trusted code does not maintain the type stack, so when returning from a trusted function the return value's type must be synthesized for the untrusted caller from the type signature.



\section{Asynchronous Checking}
\label{sec:async}

Aptos executes blocks of transactions in parallel using \emph{Block-STM}~\cite{BlockSTM}, an optimistic concurrency-control protocol in which a pool of worker threads speculatively runs transactions out of order and re-validates them against later-published writes; executions that lose a race are aborted and re-executed.
Block-STM also tracks when each transaction is committed.
If there is not enough parallelism in the block due to read-write conflicts, worker capacity remains unused.

That idle capacity can be used to run the runtime safety checks \emph{asynchronously}.
The VM performs no runtime checks but logs a compact \emph{execution trace} for each transaction, containing only control-flow shape.
After a transaction commits, a job to replay its trace is scheduled.
The replay walks the bytecode along the recorded path and runs the checks of \S\ref{sec:type} and \S\ref{sec:ability}.
Transaction commit latency is unaffected: replay runs only after the commit, proceeding in parallel with the rest of the block's execution.
A violation found during replay aborts the block before its writes are persisted on-chain.

Asynchronous checks run on committed transactions only, so speculative executions do not pay their cost.
They fail only when a static-verifier bug is being exploited, so an attacker cannot consistently slow the network by triggering such failures.


\section{Performance}
\label{sec:perf}

Our real-world benchmark, derived from Decibel~\cite{Decibel}, simulates a perpetual-futures DEX with 100 markets, driven by a weighted mix of oracle price updates, bulk market-maker orders, and retail taker orders.
It is executed by Block-STM~\cite{BlockSTM} on a 60-vCPU GCP instance with 235\,GB of RAM; worker pool varies from 2 to 32 threads.
Each reported number is the mean of three runs; run-to-run spread is below 1.3\%.

\begin{table}[t]
\caption{Throughput (txns/sec, higher is better) under type and ability checking.
Slowdown vs.\ the no-checks baseline in parentheses; ``waiver'' skips trusted framework code (\S\ref{sec:trusted}); ``asynchronous'' defers checks to trace replay (\S\ref{sec:async}).}
\label{tab:perf:type}
\centering
\scriptsize
\renewcommand{\arraystretch}{0.9}
\setlength{\tabcolsep}{2pt}
\begin{tabular}{rrrrrr}
\toprule
        & baseline & \multicolumn{2}{c}{synchronous} & \multicolumn{2}{c}{asynchronous} \\
\cmidrule(lr){3-4}\cmidrule(lr){5-6}
workers & & no-waiver & waiver & no-waiver & waiver \\
\midrule
2  & 864  & 611 ($-$29.3\%)  & 739 ($-$14.5\%)  & 513 ($-$40.6\%)  & 624 ($-$27.7\%) \\
4  & 1427 & 1026 ($-$28.1\%) & 1236 ($-$13.4\%) & 914 ($-$35.9\%)  & 1088 ($-$23.7\%) \\
8  & 1915 & 1350 ($-$29.5\%) & 1620 ($-$15.4\%) & 1427 ($-$25.5\%) & 1598 ($-$16.5\%) \\
16 & 2019 & 1427 ($-$29.3\%) & 1691 ($-$16.2\%) & 1690 ($-$16.3\%) & 1799 ($-$10.9\%) \\
32 & 1904 & 1363 ($-$28.4\%) & 1621 ($-$14.9\%) & 1633 ($-$14.2\%) & 1716 ($-$\phantom{0}9.9\%) \\
\bottomrule
\end{tabular}
\end{table}

\begin{table}[t]
\caption{Throughput (txns/sec) with isolated runtime reference safety checking.}
\label{tab:perf:ref}
\centering
\scriptsize
\renewcommand{\arraystretch}{0.9}
\setlength{\tabcolsep}{2pt}
\begin{tabular}{rrrr}
\toprule
workers & ref checks off & ref checks on & slowdown \\
\midrule
2  & 858  & 630  & $-$26.5\% \\
4  & 1430 & 1054 & $-$26.3\% \\
8  & 1918 & 1370 & $-$28.6\% \\
16 & 2011 & 1450 & $-$27.9\% \\
32 & 1893 & 1375 & $-$27.4\% \\
\bottomrule
\end{tabular}
\end{table}

Type and ability checks share their implementation, so Table~\ref{tab:perf:type} reports their combined cost (without reference checks).
The combined checks cost 28--29\% of throughput, uniformly across worker counts.
The trusted-code waiver (\S\ref{sec:trusted}) roughly halves this to 13--16\%, as about two-thirds of executed instructions on this workload fall into trusted framework code.
The asynchronous mode (\S\ref{sec:async}) trades a small tracing overhead for deferred checking on idle workers: below 8 workers it is slower than synchronous (no idle threads), reaches parity at 8, and at 16--32 brings the cost to 10--11\%, the preferred configuration at production concurrency.

The reference checker is a fully separate mechanism with its own shadow state; Table~\ref{tab:perf:ref} reports it in isolation (type and ability checks disabled): a 26--29\% slowdown, uniform across worker counts.
Reference checking is fully functional but not yet optimized, and does not yet support the trusted-code waiver or asynchronous deferral; it is therefore not yet live in production.
The type-check numbers above set a reasonable expectation for what those mechanisms could recover.


\section{Related Work and Conclusion}
\label{sec:concl}

Static bytecode verification at load time is standard for typed managed VMs: the JVM~\cite{jvm} and CLR~\cite{clr} check well-formedness, typing, and stack discipline before execution, and sibling Move VMs share Move's verifier.
None of these adds a \emph{second}, runtime layer of safety checks: once a module is admitted, its dynamic semantics is trusted.
Dynamically typed languages such as JavaScript and Python sit at the opposite end, performing the runtime type checks of \S\ref{sec:type} but lacking the \emph{first} static layer.
The EVM~\cite{ethereum,kevm} and Solana's SVM form a third regime: their untyped bytecode admits neither a static type verifier nor a runtime type discipline.
Aptos's pairing of a static verifier with a complementary runtime layer is thus unique among production VMs.

The runtime reference safety checks (RRSC) we describe are novel.
The closest related work, Tree Borrows and Stacked Borrows~\cite{stackedborrows,treeborrows} (TB/SB), clarifies Rust's unsafe aliasing rules, but does not share RRSC's requirement of no false positives over the static verifier.
TB/SB also tolerate two or more orders of magnitude of overhead, not being intended for production or large fuzzing workloads; unlike RRSC they maintain provenance genealogies among references and are not function-modular.
Also, Move and Rust's reference semantics differ (\S\ref{sec:move}).
As future work, the dynamic semantics of the reference checker is also interesting as a model for what reference safety means, since static borrow analysis over-approximates in a method-dependent way and no canonical operational semantics exists; we plan to fully formalize this.

Deferring runtime checks off the execution path and replaying a recorded trace later is a well-known idea: decoupled dynamic analysis~\cite{aftersight,shadowreplica,straighttaint} and offline runtime verification~\cite{rvsurvey} move expensive monitoring onto spare cores.
Our asynchronous checks differ in that the parallel blockchain execution model itself supplies the idle capacity (from workers waiting on read-write conflicts) and runs all checks online, so an offending transaction is detected automatically.

To conclude, we argue the need for runtime safety checks, motivated by real-world bugs, as a second line of defense to the static verifier; they also serve as a semantic, precise oracle for fuzzing the verifier itself.
We presented a novel runtime reference safety checker (yet to be launched in production) and two techniques for reducing the cost of runtime checks in a blockchain setting, with the type and ability checks (in production) introducing about 10\% overhead on Aptos workloads.

\begin{credits}
\vspace{0.0ex}
\paragraph{Acknowledgement.} Aptos auditor firm OtterSec under lead of Robert Chen highly influenced this work.
\vspace{-0.5ex}
\paragraph{Rights retention.} For the purpose of open access, the authors have applied a Creative Commons Attribution (CC BY) 4.0 license to any Author Accepted Manuscript version arising from this submission.
\vspace{-0.5ex}
\paragraph{\discintname}
All authors are employees of Aptos Labs and may hold equity in the
company. The work described is implemented in the Aptos blockchain.
\end{credits}

\bibliographystyle{splncs04}
\bibliography{biblio}

\end{document}